**Domain Nucleation and Hysteresis Loop Shape in Piezoresponse Force Spectroscopy**

Anna N. Morozovska*

V. Lashkaryov Institute of Semiconductor Physics, National Academy of Science of Ukraine,

41, pr. Nauki, 03028 Kiev, Ukraine

Eugene A. Eliseev

Institute for Problems of Materials Science, National Academy of Science of Ukraine,

3, Krjijanovskogo, 03142 Kiev, Ukraine

Sergei V. Kalinin,†

Condensed Matter Sciences Division, Oak Ridge National Laboratory, Oak Ridge, TN 37831

## Abstract

Electromechanical hysteresis loop measurements in Piezoresponse Force Microscopy (Piezoresponse Force Spectroscopy) have emerged as a powerful technique for probing ferroelectric switching behavior on the nanoscale. Interpretation of PFS data requires the relationship between the domain parameters and PFM signal to be established. Here, we analyze the switching process using a modified point charge model for the electric field of the tip. The charge value and position are selected so that its electric field isopotential surface reproduces the tip with definite radius of curvature. Using linear theory of elasticity the relationship between the sizes of semiellipsoidal domain and PFM signal has been derived. The role of domain nucleation on piezoresponse hysteresis loop is established.

* Corresponding author, morozo@i.com.ua

† Corresponding author, sergei2@ornl.gov

Polarization switching in ferroelectric materials and devices is the functional basis for applications such as non-volatile ferroelectric memories (FeRAM) and ferroelectric data storage. Piezoresponse Force Spectroscopy (PFS) is rapidly emerging as a powerful technique for probing polarization switching mechanism on the nanoscale, providing an ultimate tool to study physical properties of low-dimensional ferroelectrics. A number of studies relating PFS data to crystallographic orientation,[1,2] materials composition,[3] polarization pinning at interfaces,[4,5] etc. have been reported. Measured in PFS is the local electromechanical hysteresis loops that represent the bias dependence of local electromechanical response. The latter, in turn, is related to the geometric parameters of the domain formed below the tip. The domain switching and electromechanical detection are performed simultaneously, and thus the PFS hysteresis loop shape is determined by the convolution of the signal generation volume and the domain shape. Interpretation of the PFS data requires self-consistent solution of two problems, (a) domain nucleation and growth under the tip and (b) relating the domain parameters to the electromechanical response signal, for a given tip geometry. Here we analyze domain switching process using a point charge model and develop the analytical formulae relating domain parameters and PFM signal.

The polarization switching is induced by a tip of radius $R_0$ and potential $U$ located in the dielectric medium with permittivity $\varepsilon_e$ (e.g. air, $\varepsilon_e = 1$, or water meniscus formed at the tip-surface junction, $\varepsilon_e = 81$)[6,7,8] in contact with the surface of the transversally isotropic material, with dielectric permittivity values $\varepsilon_c$ and $\varepsilon_a$ along and perpendicular the symmetry axis. The tip-surface system is typically approximated using simple point-charge or sphere-plane models.[8,9,10,11,12] In the point charge models, effective charge magnitude, $Q$, and charge surface-separation, $d$, are subject to multiple uncertainties. Typically, the capacitive approximation, distance $d = R_0$ and charge $Q = 4\pi\varepsilon_0\varepsilon_e R_0 U\left(\kappa+\varepsilon_e\right)/\left(\kappa-\varepsilon_e\right)\ln\left(\left(\kappa+\varepsilon_e\right)/2\varepsilon_e\right)$, corresponding to capacitance of the sphere in contact with the surface, is used. However, this model does not reproduce potential behavior in the vicinity of the tip-surface junction. More rigorous sphere-plane models usually require summation over many ($N$ = 100 – 1000) image charges to adequately describe field concentration in the tip-surface junction. Here, we develop the modified point charge model, in which charge parameters ($Q, d$) are chosen such that (a) potential on the surface is equal to the tip bias and (b) the radius of curvature of isopotential surface is equal to $R_0$ in the point of contact [Fig. 1(a,b)].

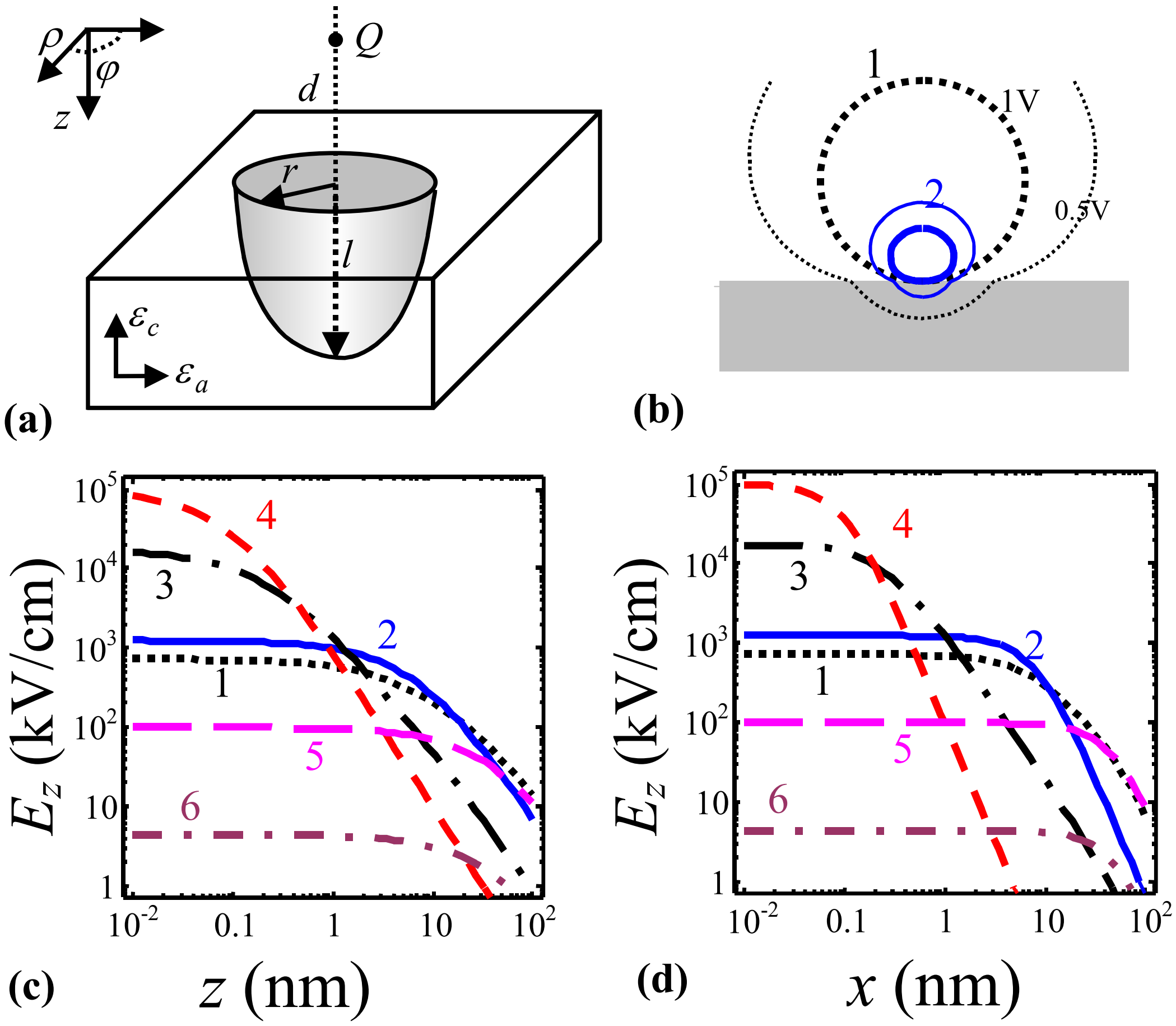

**Fig. 1.** (a) Model schematics (b) Isopotential surfaces for sphere-plane (curve 1) and effective point charge (curve 2) models calculated for $R_0 = 50\,nm$, $U = 1$ V, $\kappa = 500$ and $\varepsilon_e = 81$. Normal component of the electric field in (c) $z$ direction and (d) in radial direction for $\varepsilon_e = 81$ (1 - sphere-plane model, 2 – effective point charge, 5 – capacitance approximation) and $\varepsilon_e = 1$ (3 - sphere-plane model, 4 – effective point charge, 6 – capacitance approximation).

For transversally isotropic dielectric material, the potential above the surface is

$$V_s(\rho, z) = \frac{Q}{2\pi\varepsilon_0(\kappa+\varepsilon_e)}\left(\frac{1}{\sqrt{\rho^2+(z/\gamma-d)^2}} - \frac{\kappa-\varepsilon_e}{\kappa+\varepsilon_e}\frac{1}{\sqrt{\rho^2+(z/\gamma+d)^2}}\right), \tag{1}$$

where $\rho$ and $z$ are radial and vertical coordinate, $\kappa = \sqrt{\varepsilon_c\varepsilon_a}$ is effective dielectric constant, and $\gamma = \sqrt{\varepsilon_c/\varepsilon_a}$ is dielectric anisotropy factor. From the aforementioned conditions, the charge parameters are calculated as $d = \varepsilon_e R_0/\kappa$ and $Q = 2\pi\varepsilon_0\varepsilon_e R_0 U(\kappa+\varepsilon_e)/\kappa$. Compared in Fig. 1 (b) are the isopotential surfaces calculated for the sphere-plane model and effective point charge. The electric field distribution in normal and radial directions is illustrated in Figs. 1 (c,d).

To calculate the shape of the PFM hysteresis loop, the electromechanical response induced by point charge $Q$ located at distance $d$ above the domain of length $l$ and radius $r$ is required [Fig. 1(a)]. Here, we employ the linearized theory by Felten et al.[7] to calculate electromechanical response. The displacement vector $u_i(\mathbf{x})$ at position $\mathbf{x}$ is

$$u_i(\mathbf{x}) = \int_0^{\infty} d\xi_3 \int_{-\infty}^{\infty} d\xi_2 \int_{-\infty}^{\infty} d\xi_1 e_{kjl} E_k(\xi) \frac{\partial G_{ij}(\mathbf{x},\xi)}{\partial \xi_l} \tag{2}$$

where $\xi$ is the coordinate system related to the material, $e_{kjl}$ are the piezoelectric coefficients ($e_{kij} = d_{klm} c_{lmij}$, where $d_{klm}$ are strain piezoelectric coefficients and $c_{lmij}$ are elastic stiffness) and the Einstein summation convention is used. $E_k(\xi)$ is the electric field produced by the charge, i.e. probed volume. For typical ferroelectric perovskites, the symmetry of the elastic properties can be approximated as cubic (anisotropy of elastic properties is much smaller then that of dielectric and piezoelectric properties) and therefore isotropic approximation is used for the Green's function, $G_{ij}(\mathbf{x},\xi)$.[13] The applicability of this decoupled approximation for PFM is analyzed in detail elsewhere.[14, 15]

Integration of Eq. (2) for $z = 0, \rho = 0$ over semi-ellipsoidal domain with semi-axes $r$ and $l$ [Fig. 1(a)] yields $u_3^i(0) = u_3(0) - 2\tilde{u}_3(0)$, where $\tilde{u}_3(0)$ is response from semi-ellipsoidal domain and $u_3(0)$ is response from semi-infinite material corresponding to the initial state of the ferroelectric. After lengthy integration, Eq. (2) yields

$$u_3^i = \frac{Q}{2\pi\varepsilon_0(\varepsilon_e + \kappa)} \frac{1}{d} \frac{1+\nu}{Y} \left(e_{31} g_1(\gamma) + e_{15} g_2(\gamma) + e_{33} g_3(\gamma)\right), \tag{3}$$

where $Y$ is Young's modulus, $\nu$ is Poisson ratio and $g_i(\gamma) = f_i(\gamma) - 2w_i(\gamma)$. Functions $w_i(\gamma) = 0$ in the initial and $w_i(\gamma) = f_i(\gamma)$ in the final state of the switching process. The functions $f_i(\gamma)$ are given by $f_1(\gamma) = \left(\frac{\gamma}{(1+\gamma)^2} - \frac{(1-2\nu)}{(1+\gamma)}\right)$, $f_2(\gamma) = -\frac{2\gamma^2}{(1+\gamma)^2}$ , $f_3(\gamma) = -\left(\frac{\gamma}{(1+\gamma)^2} + \frac{(1-2\nu)}{1+\gamma}\right)$ and define the contributions of different piezoelectric constant to PFM response in the initial and final states of switching process.[16].

The functions $w_i(\gamma)$ are dependent on the domain sizes and can be reduced to the integral representations

$$w_1(\gamma,s,\delta)=\int_0^1 d\xi\frac{s^2\cdot h_1\left(\xi/\sqrt{s^2+\left(1-s^2\right)\xi^2}\right)}{\left(s^2+\left(1^2-s^2\right)\xi^2\right)\sqrt{(\delta+\xi)^2+\gamma^2 s^2\left(1-\xi^2\right)}},\tag{4}$$

$$w_2(\gamma,s,\delta)=\int_0^1 d\xi\frac{s^2 h_2\left(\xi/\sqrt{s^2+\left(1-s^2\right)\xi^2}\right)}{\left(s^2+\left(1^2-s^2\right)\xi^2\right)^{3/2}}\left(\frac{\delta+\xi}{\sqrt{(\delta+\xi)^2+\gamma^2 s^2\left(1-\xi^2\right)}}-1\right),\tag{5}$$

$$w_3(\gamma,s,\delta)=\int_0^1 d\xi\frac{s^2\cdot h_3\left(\xi/\sqrt{s^2+\left(1-s^2\right)\xi^2}\right)}{\left(s^2+\left(1^2-s^2\right)\xi^2\right)\sqrt{(\delta+\xi)^2+\gamma^2 s^2\left(1-\xi^2\right)}},\tag{6}$$

where $h_1(\zeta)=-\zeta+3\zeta^3-(1-2v)\zeta$, $h_2(\zeta)=6\zeta^2$, and $h_3(\zeta)=\zeta-3\zeta^3-(1-2v)\zeta$. Domain geometry in Eqs. (4-6) is described by dimensionless domain aspect ratio $s=r/l$ and charge-surface separation $\delta=\gamma d/l$ . For more complex tip geometries, Eqs. (3-6) can be summed over corresponding image charge series (Appendix A).

Note that due to the non-zero charge-surface separation, the electric field on the surface is finite, resulting in critical tip bias required for the nucleation of ferroelectric domain. To determine the domain size as a function of tip bias, the former is calculated for semi-infinite ferroelectric material using thermodynamic formalism developed by Morozovska and Eliseev.[17] The Pade approximations for the individual terms in free energy $\Phi(r,l)=\Phi_S(r,l)+\Phi_U(r,l)+\Phi_D(r,l)$ for the semi-ellipsoidal domain derived for ferroelectrics-semiconductors allowing for Debye screening and uncompensated surface charges are listed in Appendix B. We consider the domain wall surface energy $\Phi_S(r,l)$ , interaction energy $\Phi_U(r,l)$ and depolarization field energy $\Phi_D(r,l)=\Phi_{DL}(r,l)+\Phi_{DS}(r,l)$ allowing for the Landauer contribution $\Phi_{DL}(r,l)$ and the energy $\Phi_{DS}(r,l)$ created by the surface charges located on the domain surface. The latter has not been considered previously,[18,10,19] and its inclusion can significantly affect the predicted behavior.

The free energy evolution with voltage is shown in Fig.2. The equilibrium domain size can be determined from the minimum of the $\Phi(r,l)$ surface. Activation energy for nucleation and critical domain size are determined from the saddle point on the $\Phi(r,l)$ surface. Finally, nucleation bias corresponds to the condition when energy corresponding to the minimum on the $\Phi(r,l)$ surface becomes negative, i.e. domain becomes thermodynamically stable.

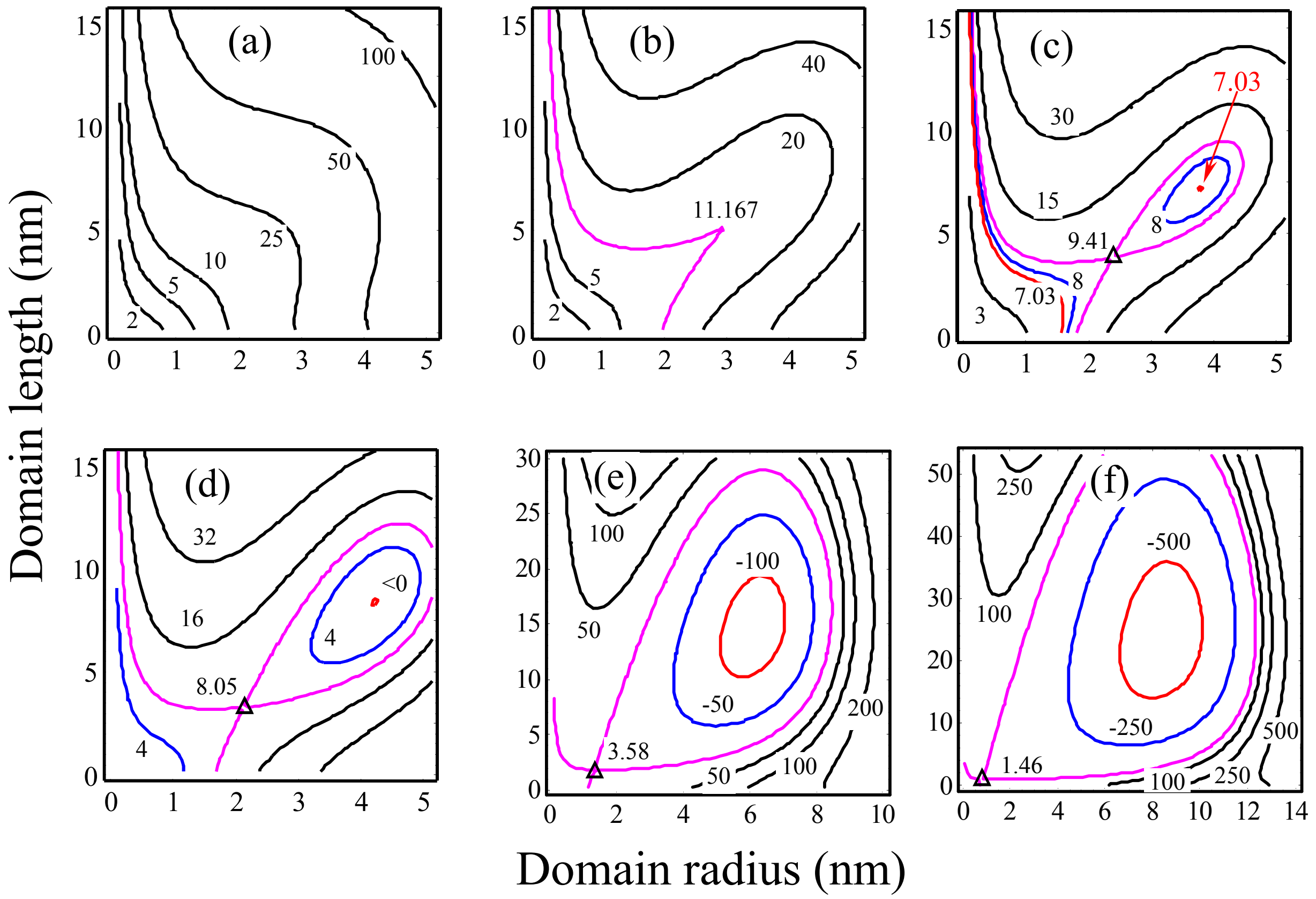

**Fig. 2.** Contour plots of the free energy surface under the voltage increase: (a) domain is absent ($U = 2$ V); (b) critical point ($U = 2.344$ V); local minimum became to appear; (c) saddle point and metastable domain appears ($U = 2.4$ V); (d) transition point ($U = 2.467$ V); stable domain appears; (e, f) stable domains growth ($U = 3; 4$ V). Figures near the contours are free energy values in eV. Triangles denote saddle point (nuclei sizes). Material parameters: Debye screening radius $R_d = 500\,nm$, $P_S \approx 0.5\ C/m^2$, $\psi_S \approx 150\,mJ/m^2$, $\varepsilon_a \approx 515$, $\varepsilon_c = 500$ correspond to the PZT6B solid solution and tip-surface characteristics: $\varepsilon_e = 81$, $R_0 = 50nm$, tip touches the sample; uncompensated surface charges density $\sigma_S = -P_S$.

Shown in Figs.3 are the activation energy for nucleation (a) and critical domain sizes calculated by exact image charges series (b) and modified point charge model (c) for different screening conditions on the surface (compare dotted and dashed curves in Fig. 3). The bias dependence of the equilibrium domain energy is shown in Fig. 3 (e) and equilibrium domain sizes in exact sphere plane and modified point charge models are shown in Figs. 3 (e,f) correspondingly.

Note that the critical domain shape is close to the semi-spherical independently on the adopted model, whereas equilibrium domain is always prolate [compare Fig.3 (b,c) with Fig.3 (e,f)].

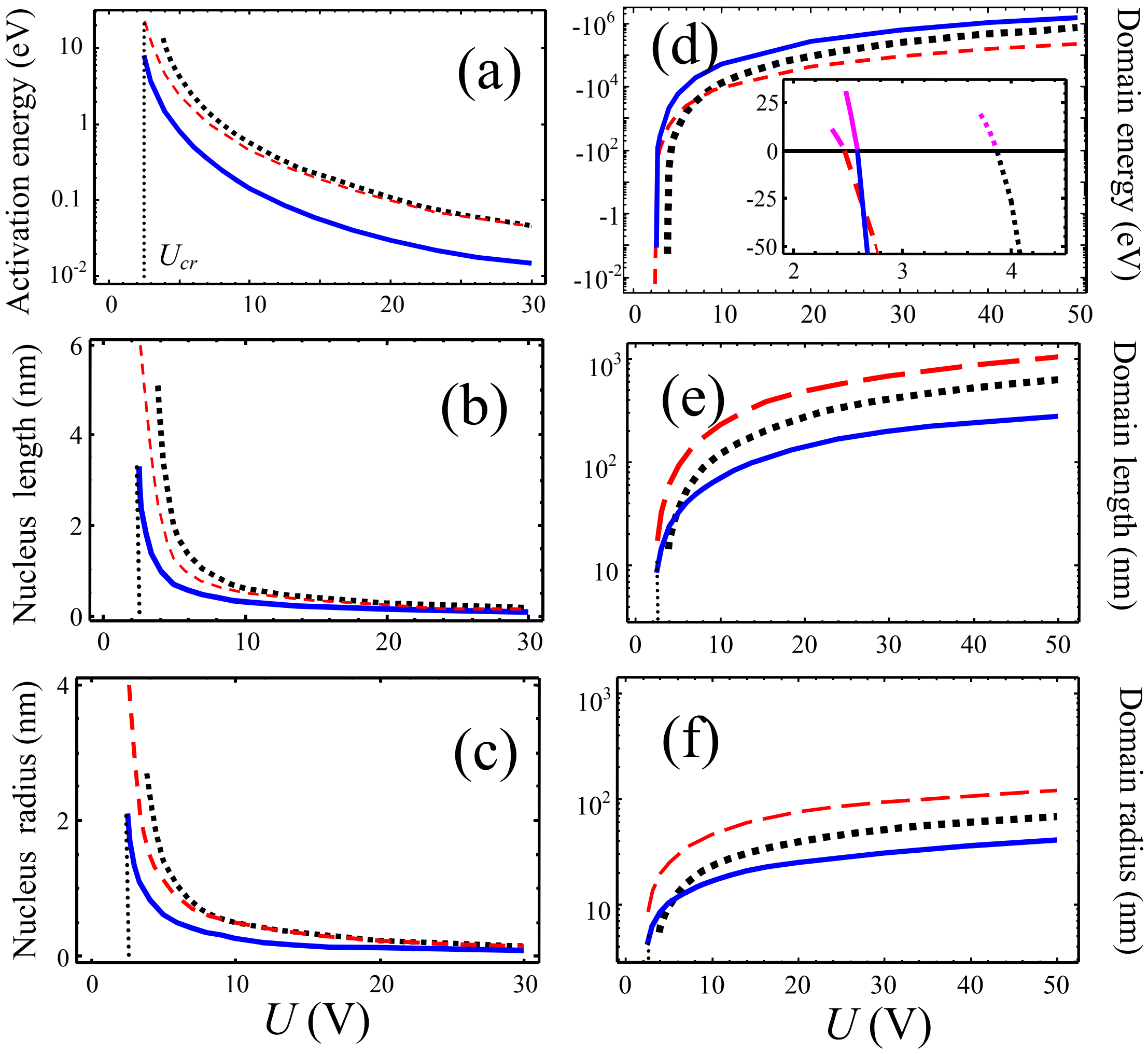

**Fig. 3.** (a) Activation energy (eV) and nucleus sizes (b,c), (d) domain energy (eV) and equilibrium domain sizes (nm) (e,f) vs. voltage $U$ (V) calculated for PZT6B. Solid curves represent modified point charge approximation of the tip; dotted ones correspond to the exact series for sphere-tip interaction energy. Dashed curves represent the model without surface charges depolarization energy $\Phi_{DS}(r,l)$. Material parameters are given in Fig.2.

From Fig. 3, domain formation is impossible below certain nucleation bias, $U_{cr}$. The critical domain size rapidly decreases with bias. Note, that the effect of depolarization field due to the surface termination of the domain decreases the equilibrium domain sizes up to several times for PZT, however, it does not affect the activation energy for voltages $U \geq 10\,V$ (compare Fig. 3 (a) with Fig. 3 (d) or (f)).

Assuming that characteristic time for the nucleation is $\tau = \tau_0 \exp(E_a / k_B T)$ and attempt time $\tau_0 = 10^{-13}\,s$, the thermal activation of domain nucleation in the PFM experiment requires the activation barrier below 0.7 - 0.8eV ($\tau = 10^{-3} - 1$ s) corresponding to the tip voltages $U = 5 ... 8\,V$. Interestingly, in capacitive approximation the activation energy is very high (barrier about 200 eV for $U_{cr} \approx 10\,V$) making the thermal activation of domain nucleation process impossible. The sphere-plane and charge models developed here allow for field concentration in the tip-surface junction, and thus are suited for description of domain nucleation in PFS. Note that applicability of Landauer model requires critical domain size to be larger then several correlation lengths. The correlation length cannot be smaller than several lattice constants, thus the nucleation barrier disappears at $U \geq 15\,V$.

To calculate the thermodynamic hysteresis loop shape from the bias dependence of domain size and Eqs. (4)-(6), we assume that domain evolution follows the equilibrium domain size on the forward branch of the hysteresis loop. On reverse branch of hysteresis loop the domain does not shrink. Rather, domain wall is pinned by lattice and defect, resulting in the bias dependence of domain radius shown in Fig. 4 (a)[20]. Corresponding piezoelectric loop is shown in Fig 4 (b).

It is clear from Figs.4c,d, that the modified point charge model gives the thinner loop that saturates more quickly than the exact series for sphere-tip interaction energy and moreover quicker than capacitance approximation. This can be explained taking into account that the distance $d$ between the effective point charge $Q$ and the sample surface is smaller in $\kappa/\varepsilon_e \approx 6$ times than the first ones from the image charges caused by the tip with curvature $R_0$.

Numerically, the results obtained within modified point charge approximation of the tip can be approximated by $d_{33}^{\mathrm{eff}} = 64\left(1 - \sqrt{3.3/U}\right)$ in the entire region $U \geq U_{cr}$, where $U_{cr} = 2.49\,V$. Circles in part (b) is calculations in 1D model [1] $d_{33}^{\mathrm{eff}} = 64(1 - 3.3/U)$ (curve 1) and $d_{33}^{\mathrm{eff}} = 51.2(1 - 4/U)$ (curve 2). Keeping in mind, that the bulk value of $d_{33}^{\mathrm{eff}}$ is 64pm/V, the point charge approximation of the tip is perfectly described by the law $d_{33}^{\mathrm{eff}} = d_{\infty}\left(1 - \sqrt{U_0/U}\right)$. The difference with the one $d_{33}^{\mathrm{eff}} = d_{\infty}(1 - U_0/U)$ obtained within the framework of 1D model could be related to the dimension of the problem.

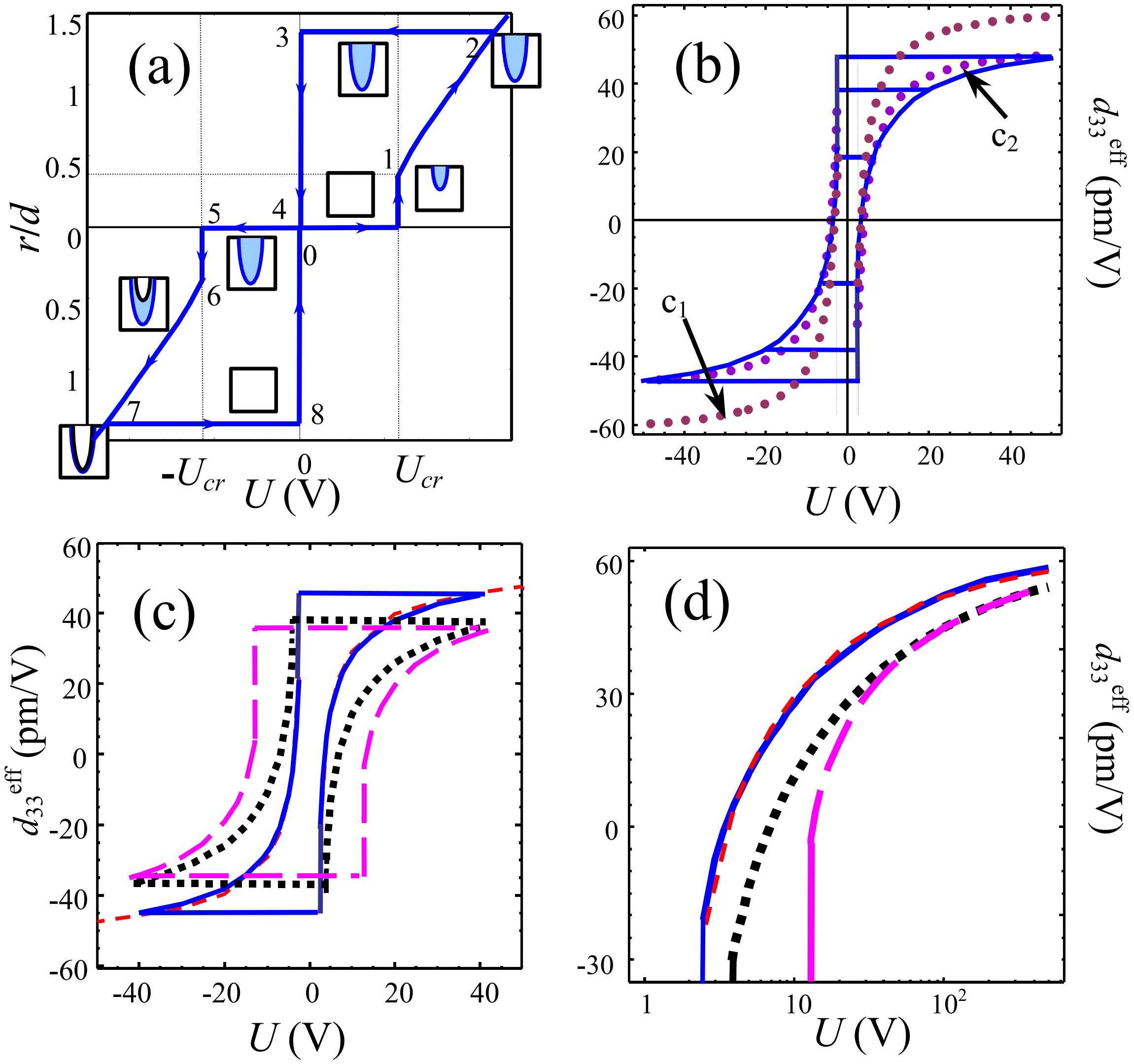

**Fig. 4** Domain radius (a) and piezoelectric response (b, c, d) as the function of applied voltage for PZT6B. Solid curves represent modified point charge approximation of the tip; dotted ones correspond to the exact series for sphere-tip interaction energy; long-dashed curves is the capacitance approximation; short-dashed ones represent the model without surface charges depolarization energy $\Phi_{DS}(r,l)$. Circles in part (b) is fitting $d_{33}^{\text{eff}} = 64(1-3.3/U)$ (curve $c_1$) and $d_{33}^{\text{eff}} = 51.2(1-4/U)$ (curve $c_2$). Upper and bottom curves correspond to the different orientations of the spontaneous polarization (sign of tensor *e*). Material parameters and tip-surface characteristics are given in Fig.2.; $e_{33} = 7.4$, $e_{31} = -0.96$ and $e_{15} = 4.8$ C/m$^2$.

In the thermodynamic models considered above, the nucleation bias is relatively small compared to the bias necessary for the saturation of response. Note that the hysteresis loops saturate relatively slow, when the response is within 90% of saturated value for domain sizes are $r \approx 110\,nm$ and $l \geq 500\ nm$ for charge surface separation of 8 nm. We ascribe this behavior to the fact that for ferroelectric materials with $\gamma < 1$, the field is concentrated primarily in the surface region. At the same time, domains usually adopt prolate geometry ($l >> r$). Hence, only the part of the domain close to the surface contributes to the PFM signal, and domain radius is the dominant length scale determining the PFS response. Relatively weak dependence of domain radius, *r*, on voltage, *U,* explains the slow saturation of the response. In realistic material, the predictions of the model considered above will be mediated by two factors. First, the domain size is likely to be limited by the kinetics of domain wall motion; in this case the both domain length and radius will grow slower then predicted by thermodynamic model. However, due to the rapid decay of the field in z-direction, this will result in larger r/l ratio. Secondly, the diffusion of charged species on the surface can result in rapid broadening of the domain in r-direction. Given that only the part of the surface in contact with the tip results in cantilever deflection (i.e. electrical radius is much larger then mechanical radius), this may result in rapid saturation of the hysteresis loop.

To summarize, the hysteresis loop formation mechanism in PFM is analyzed using semi-ellipsoid approximation for domain geometry and improved point charge model for the tip. The point charge parameters are selected to reproduce tip potential and radius of curvature near the surface. The bias dependence of domain parameters is shown to have pronounced nucleation stage. The relationship between domain geometry and PFM signal is established, providing the loop shape. Our consideration within the framework of the modified point charge model leads to realistic values of the activation energy (less than 0.1eV for *U*>10V).

Research supported by Oak Ridge National Laboratory, managed by UT-Battelle, LLC, for the U.S. Department of Energy under Contract DE-AC05-00OR22725.

**Appendix A**

Under the conditions, $s < 1$ and $\delta << 1$, approximate expressions for $w_i$, are found as

$$w_1 \approx f_1(\gamma)\left(1 - \frac{\delta\left(\delta + \tilde{\xi}\right)}{\left(\delta + \tilde{\xi}\right)^2 + \gamma^2 s^2\left(1 - \tilde{\xi}^2\right)}\right), \tag{A.1}$$

$$w_2 \approx f_2(\gamma)\left(1 - \frac{\delta}{\sqrt{\tilde{\xi}^2 + \gamma^2 s^2\left(1 - \tilde{\xi}^2\right)}}\left(1 + \frac{\tilde{\xi}}{\sqrt{\tilde{\xi}^2 + \gamma^2 s^2\left(1 - \tilde{\xi}^2\right)}}\right)\right), \tag{A.2}$$

$$w_3 \approx f_3(\gamma)\left(1-\frac{\delta\left(\delta+\tilde{\xi}\right)}{\left(\delta+\tilde{\xi}\right)^2+\gamma^2 s^2\left(1-\tilde{\xi}^2\right)}\right). \tag{A.3}$$

where $\tilde{\xi}(\gamma,s)=s/(s+\gamma)$. Exact solutions (solid curves) for $w_i$ are compared with approximation (dashed curves) for different values dielectric anisotropy values $\gamma$ in Fig. 1A.

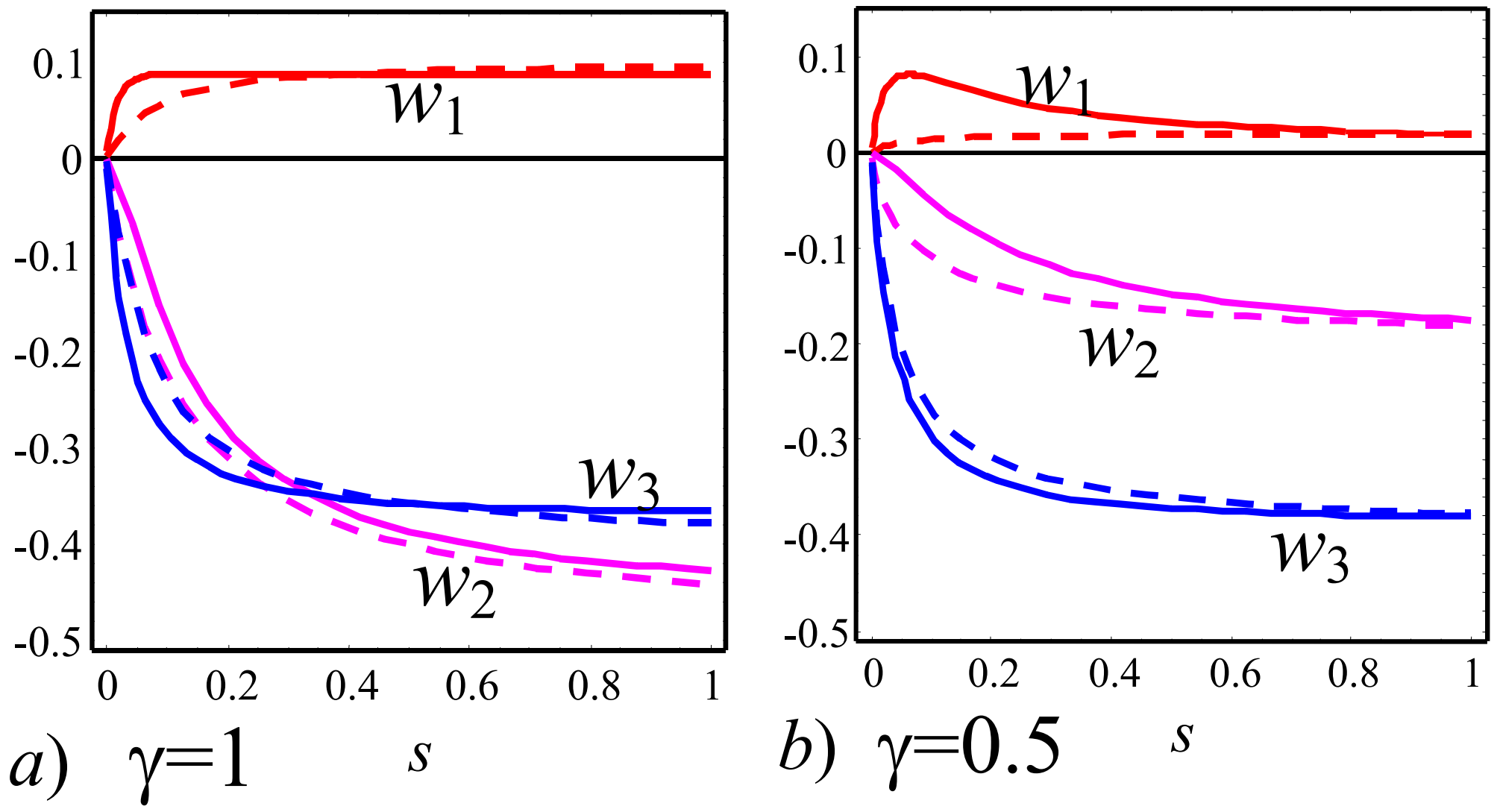

**Fig. 1A.** Comparison of exact solutions for $w_i$ (solid curves calculated from Eqs. (4)-(6)) with approximate ones (dashed curves calculated from Eqs. (A.1)-(A.3)) for $\nu = 0.35$, $\delta = 0.1$ and different dielectric anisotropy values $\gamma = 1$ (a), $\gamma = 5$ (b).

Notably, that the positive term $w_1$ (the worst comparison between exact and approximate formulae) is much smaller than the negative ones $w_2 + w_3$ that provide a reasonable agreement with exact expressions.

In the case of spherical tip that touches the sample surface Eq.(3) should be substituted by series on image charges, namely:

$$u_3^i = \frac{Q}{2\pi\varepsilon_0(\varepsilon_e+\kappa)}\frac{1+\nu}{Y}\sum_{m=0}^{\infty}\frac{q_m}{R_0-r_m}\left(e_{31}g_1(\gamma)+e_{15}g_2(\gamma)+e_{33}g_3(\gamma)\right)$$
$$g_i(\gamma)=f_i(\gamma)-2\,w_i(\gamma,s,\delta_m), \tag{A.4}$$
$$Q=4\pi\varepsilon_0\varepsilon_e U R_0,\quad q_m=\left(\frac{\kappa-\varepsilon_e}{\kappa+\varepsilon_e}\right)^m\frac{1}{m+1},\quad r_m=\frac{mR_0}{m+1},\qquad \delta_m=\frac{\gamma R_0}{(m+1)l}$$

## Appendix B

The important experimental fact should be taken into consideration. Namely, in his analysis of PFM experiment Durkan et al.[6] reported the layer of the adsorbed water located below the tip apex. Many authors [7] assume, that a water meniscus appears between the AFM tip apex and a sample surface due to the air humidity. Hereinafter we regard, that the region between the tip apex and domain surface has effective dielectric permittivity $\varepsilon_e$.

Aforementioned adsorbed water layer provides additional screening of the bond charges owing to dissocation into ions, thus this effect exists even in the absence of applied voltage. The relevance of screening mechanism to the polarization switching process depends on the relationship between its characteristic relaxation time $\tau_S$ and voltage pulse time $\tau_U$ (i.e. recording time of the domain). "Fast" screening mechanisms with $\tau_S \le \tau_U$ should be considered under the domain formation, whereas the "slow" ones with $\tau_S >> \tau_U$ could be neglected. However slow mechanisms have to be considered when applied voltage is turned off, since they affect domain stability during many days and weeks. Here, we assume surface charge density $\sigma_S$ has the form:

$$\begin{cases} \sigma_S = -P_S, & \textit{without screening charges} \\ -P_S < \sigma_S < P_S, & \textit{partial screening} \\ \sigma_S = +P_S, & \textit{full screening} \end{cases} \qquad \text{(B.1)}$$

Below we propose the Pade approximations for the components of the semi-ellipsoidal domain free energy $\Phi(r,l) = \Phi_S(r,l) + \Phi_U(r,l) + \Phi_D(r,l)$ derived for ferroelectrics-semiconductors allowing for surface charges. The question about their accuracy and related mathematical calculations are discussed in details in Appendix C.

1. Domain wall energy

The domain wall surface energy $\Phi_S(r,l)$ has the form:

$$\Phi_S(r,l) = \pi\psi_S\, l\, r\left(\frac{r}{l} + \frac{\arcsin\sqrt{1-r^2/l^2}}{\sqrt{1-r^2/l^2}}\right) \approx \frac{\pi^2\psi_S\, l\, d}{2}\left(1 + \frac{2(d/l)^2}{4+\pi(d/l)}\right) \qquad \text{(B.2)}$$

2. Interaction energy

Exact calculations of interaction energy are very complex. We develop Pade approximation for the interaction energy series on image charges (see Appendix C)

$$\Phi_U(r,l) \approx 4\pi\varepsilon_0\varepsilon_e U R_0 \sum_{m=0}^{\infty} q_m \frac{R_d\left((\sigma_S - P_S)F_W(r,0,d-r_m) + 2P_S F_W(r,l,d-r_m)\right)}{\varepsilon_0\left((\kappa+\varepsilon_e)R_d + 2\kappa\sqrt{(d-r_m)^2 + r^2}\right)}$$

$$q_0 = 1, \quad q_m = \left(\frac{\kappa-\varepsilon_e}{\kappa+\varepsilon_e}\right)^m \frac{\sinh(\theta)}{\sinh((m+1)\theta)}, \tag{B.3}$$

$$r_0 = 0, \quad r_m = R_0 \frac{\sinh(m\theta)}{\sinh((m+1)\theta)}, \quad \cosh(\theta) = \frac{d}{R_0}$$

Here the function $F_W(r,l,z) \approx \frac{r^2}{\sqrt{r^2+z^2} + z + (l/\gamma)}$ is the Pade approximation of cumbersome exact expression obtained by Molotskii [10].

Under the typical conditions $\Delta R << R_0$ and $r < R_0$ ($R_0$ is the tip radius of curvature, $\Delta R$ is the distance between the tip apex and sample surface) the modification of Eq.(3) is valid:

$$\Phi_U(r,l) \approx \frac{R_d U C_t/\varepsilon_0}{(\kappa+\varepsilon_e)R_d + 2\kappa\sqrt{d^2+r^2}} \left(\frac{(\sigma_S - P_S)r^2}{\sqrt{r^2+d^2}+d} + \frac{2P_S r^2}{\sqrt{r^2+d^2}+d + (l/\gamma)}\right) \tag{B.4}$$

Hereinafter the distance from the equivalent charge $d$ and the effective tip capacity $C_t$ has been introduced. Note, that earlier Molotskii [10], Morozovska and Eliseev [11] assumed $d = R_0 + \Delta R \approx R_0$ and $C_t \approx 4\pi\varepsilon_0\varepsilon_e R_0 \sum_{m=0}^{\infty}\left(\frac{\kappa-\varepsilon_e}{\kappa+\varepsilon_e}\right)^m \frac{\sinh(\theta)}{\sinh((m+1)\theta)}$ (*a*), whereas Kalinin and Abplanalp [9] proposed the effective point charge models approach (*b*). Within the approach $d = \varepsilon_e R_0/\kappa$ and $C_t \approx 4\pi\varepsilon_0\varepsilon_e R_0 \frac{\kappa+\varepsilon_e}{2\kappa}$ respectively (the equipotential surface produced by the effective point charge located at distance $d$ from the sample touches its surface in the point (0,0,0) with curvature $R_0$; effective charge value $Q = U \cdot C_t$ gives charge potential equal to $U$ at this equipotential surface).

3. The depolarization field energy

The energy of the depolarization field is created by surface charges and bulk charges (Landauer energy [18]), i.e. $\Phi_D(r,l) = \Phi_{DL}(r,l) + \Phi_{DS}(r,l)$. Pade approximation for the Landauer energy of semi-ellipsoidal domain (allowing for Debye screening) acquires the form:

$$\Phi_{DL}(r,l) = \frac{4\pi P_S^2 r^2}{\varepsilon_0\kappa} \frac{R_d n_D(r,l)}{4n_D(r,l) + 3R_d(\gamma/l)} \tag{B.5}$$

Hereinafter the depolarization factor $n_D(r,l)=\frac{(r\gamma/l)^2}{1-(r\gamma/l)^2}\left(\frac{\mathrm{arcth}\left(\sqrt{1-(r\gamma/l)^2}\right)}{\sqrt{1-(r\gamma/l)^2}}-1\right)$ is introduced.

The energy of the depolarization field created by the surface charges $(\sigma_S-P_S)$ located on the domain face has the form:

$$\Phi_{DS}(r,l)\approx\frac{4\pi r^3 R_d}{\varepsilon_0\left(16\kappa r+3\pi R_d(\kappa+\varepsilon_e)\right)}\left((\sigma_S-P_S)^2+\frac{2P_S(\sigma_S-P_S)}{1+(l/r\gamma)}\right) \tag{B.6}$$

Let us underline that the energy (B.6), created by the surface charges was not considered in models [10, 18, 19].

## Appendix C

a) Used in (B.2) Pade approximation $\left(x+\frac{\arcsin\sqrt{1-x^2}}{\sqrt{1-x^2}}\right)\approx\frac{\pi}{2}\left(1+\frac{2x^2}{4+\pi x}\right)$ is rather well, see plots

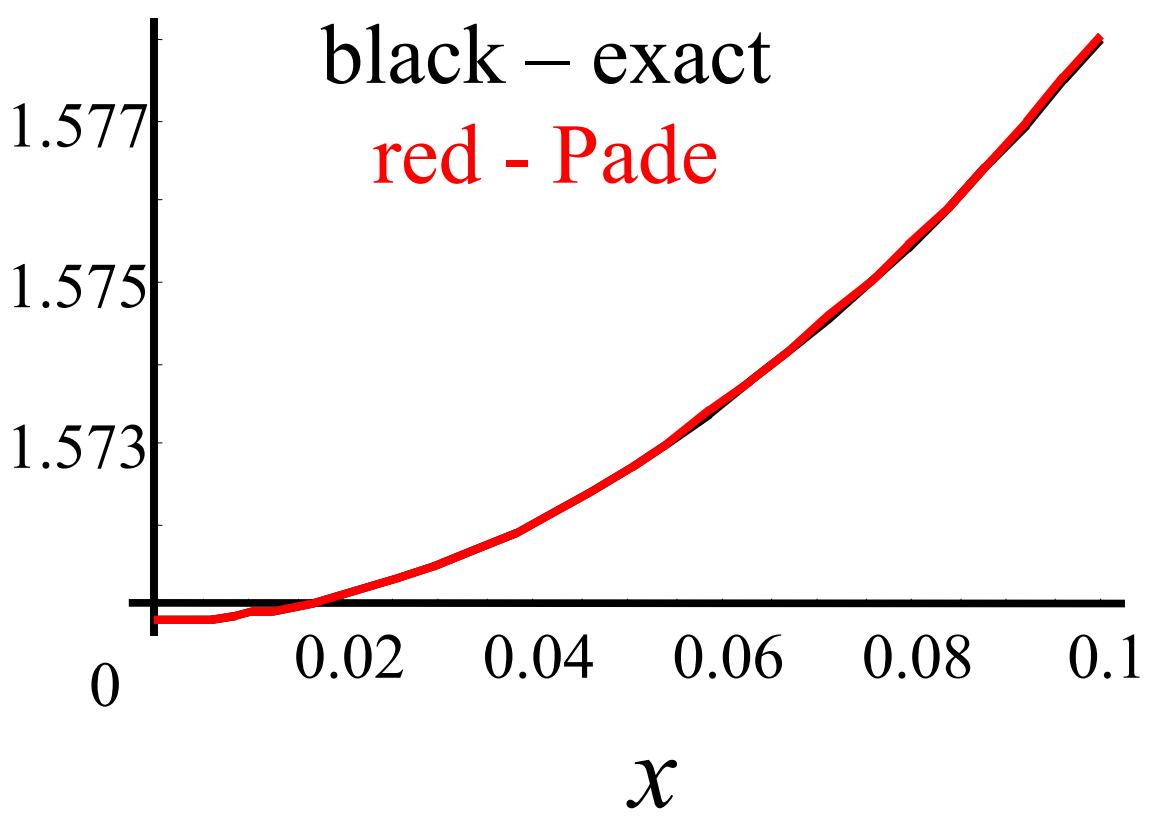

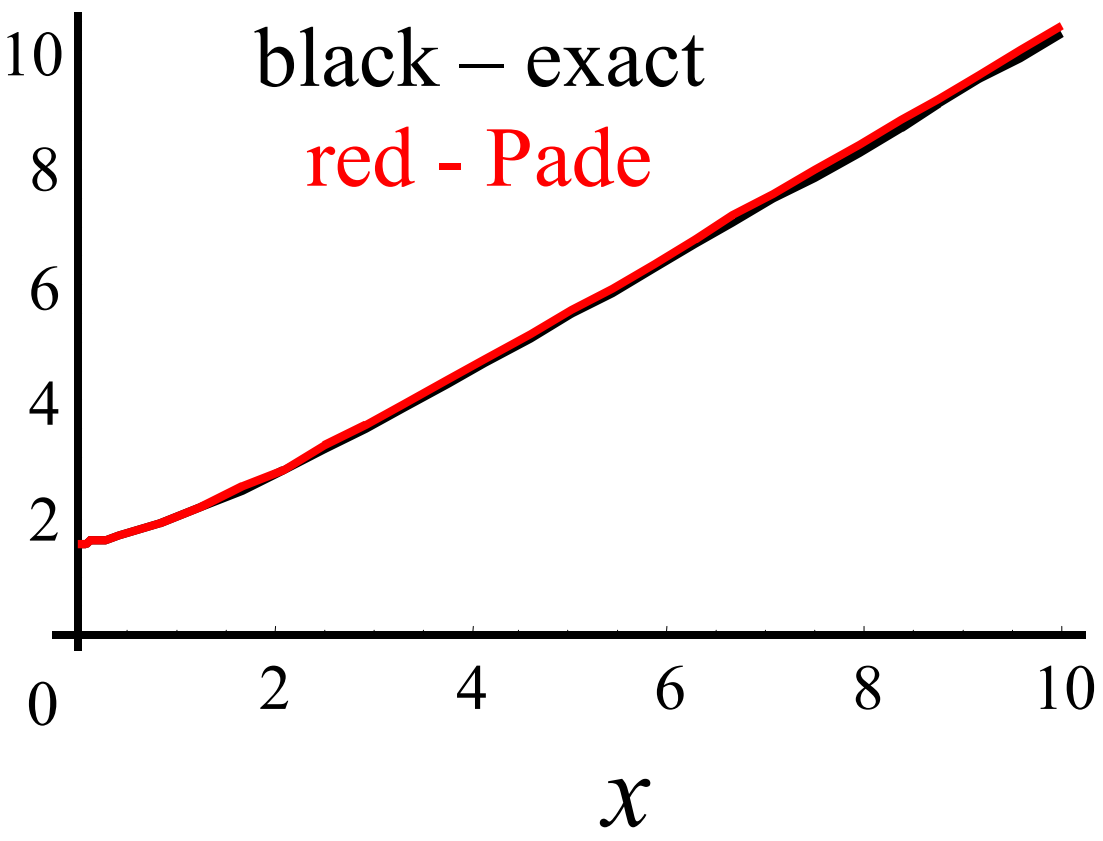

b) Exact expression for interaction energy (B.3) is the following

$$\Phi_U(r,l)\approx 4\pi\varepsilon_e U R_0\sum_{m=0}^{\infty}q_m\frac{R_d\left((\sigma_S-P_S)F_W(r,0,d-r_m)+2P_SF_W(r,l,d-r_m)\right)}{(\kappa+\varepsilon_e)R_d+2\kappa\sqrt{(d-r_m)^2+r^2}}$$

$$q_0=1,\quad q_m=\left(\frac{\kappa-\varepsilon_e}{\kappa+\varepsilon_e}\right)^m\frac{sh(\theta)}{sh((m+1)\theta)}, \tag{C.1}$$

$$r_0=0,\quad r_m=R_0\frac{sh(m\theta)}{sh((m+1)\theta)}\quad,\quad ch(\theta)=\frac{d}{R_0}$$

Exact expression for $F_W(d,l)$ is the following

$$F_W(r,l,d)=\begin{cases}\left(\dfrac{\sqrt{d^2+r^2}-(d+l/\gamma)}{(l/r\gamma)^2-1}+\dfrac{d}{\left((l/r\gamma)^2-1\right)\sqrt{1-(r\gamma/l)^2}}\times\right.\\ \left.\times\ln\left(\dfrac{d+(l/\gamma)\left(1-(r\gamma/l)^2\right)+(d+l/\gamma)\sqrt{1-(r\gamma/l)^2}}{d+\sqrt{1-(r\gamma/l)^2}\sqrt{r^2+d^2}}\right)\right) & at\ (\gamma r/l)<1\\ \left(\dfrac{\sqrt{d^2+r^2}-(d+l/\gamma)}{(l/r\gamma)^2-1}+\dfrac{d}{\left((l/r\gamma)^2-1\right)\sqrt{(r\gamma/l)^2-1}}\times\right.\\ \left.\times\left(arctg\left(\dfrac{(d+l/\gamma)\sqrt{(r\gamma/l)^2-1}}{d-(l/\gamma)\left((r\gamma/l)^2-1\right)}\right)-arctg\left(\dfrac{\sqrt{d^2+r^2}\sqrt{(r\gamma/l)^2-1}}{d}\right)\right)\right) & at\ (\gamma r/l)\gtrsim 1\end{cases}$$

(C.2)

Pade approximation $\dfrac{F_W(r,l,d)}{d}\approx\dfrac{r^2/d}{\sqrt{r^2+d^2}+d+(l/\gamma)}=\dfrac{-xy^2}{y+\left(\sqrt{1+y^2}+1\right)x}$, ( $x=(\gamma r/l)$, $y=r/d$ )

is rather well, see plots

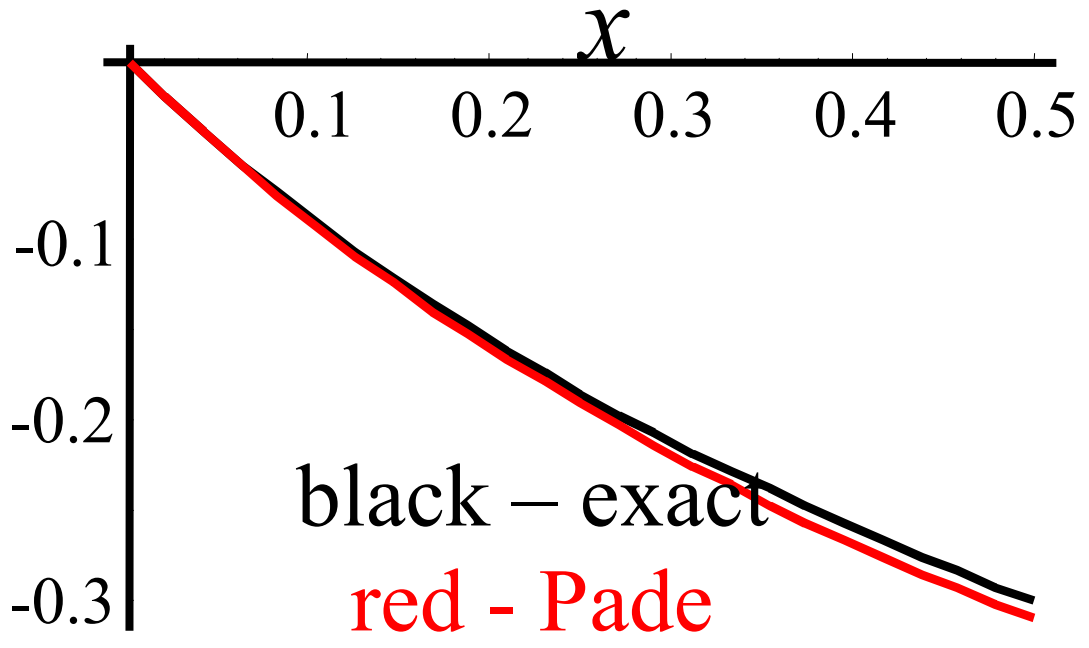

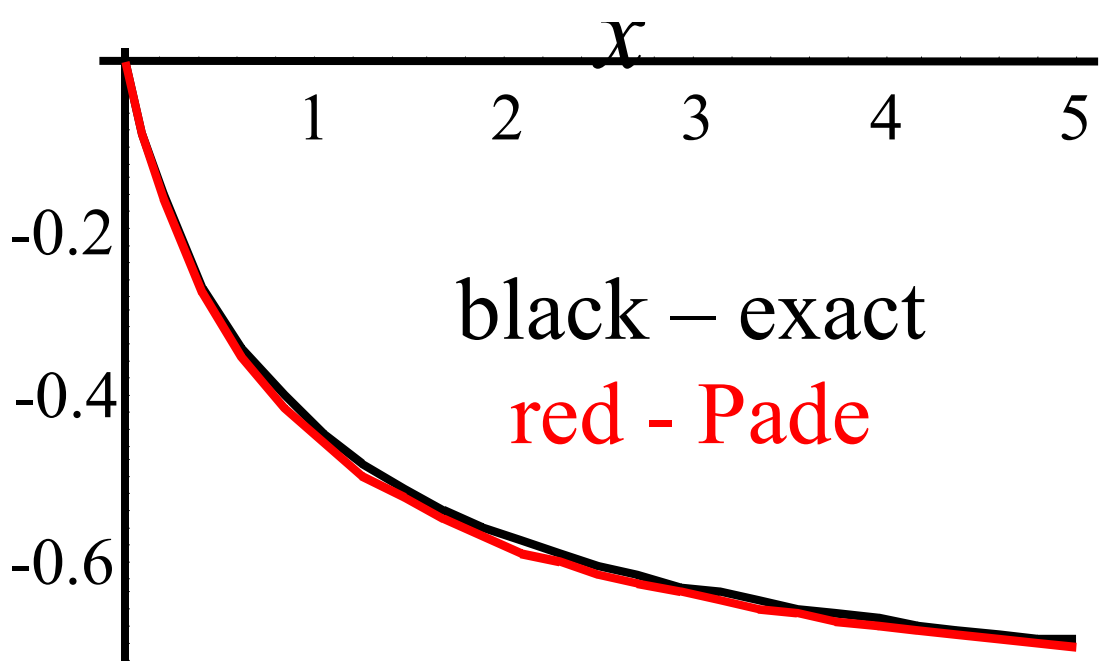

c) The exact expression for depolarization factor $n_D(r,l)$ is well known, namely

$$n_D(r,l)=\begin{cases}\dfrac{(r\gamma/l)^2}{1-(r\gamma/l)^2}\left(\dfrac{\mathrm{arcth}\left(\sqrt{1-(r\gamma/l)^2}\right)}{\sqrt{1-(r\gamma/l)^2}}-1\right), & at\quad \dfrac{l}{\gamma}\gtrsim r\\ \dfrac{(r\gamma/l)^2}{(r\gamma/l)^2-1}\left(1-\dfrac{\mathrm{arctg}\left(\sqrt{(r\gamma/l)^2-1}\right)}{\sqrt{(r\gamma/l)^2-1}}\right), & at\quad \dfrac{l}{\gamma}<r\end{cases}\qquad (C.3)$$

The proposed approximation for $n_D(x)\approx\dfrac{\sqrt{c_0^2+x^2}-c_0}{x+\pi/2}$ ( $x=(r\gamma/l)$ ) is good even for derivatives at $x>0.05$

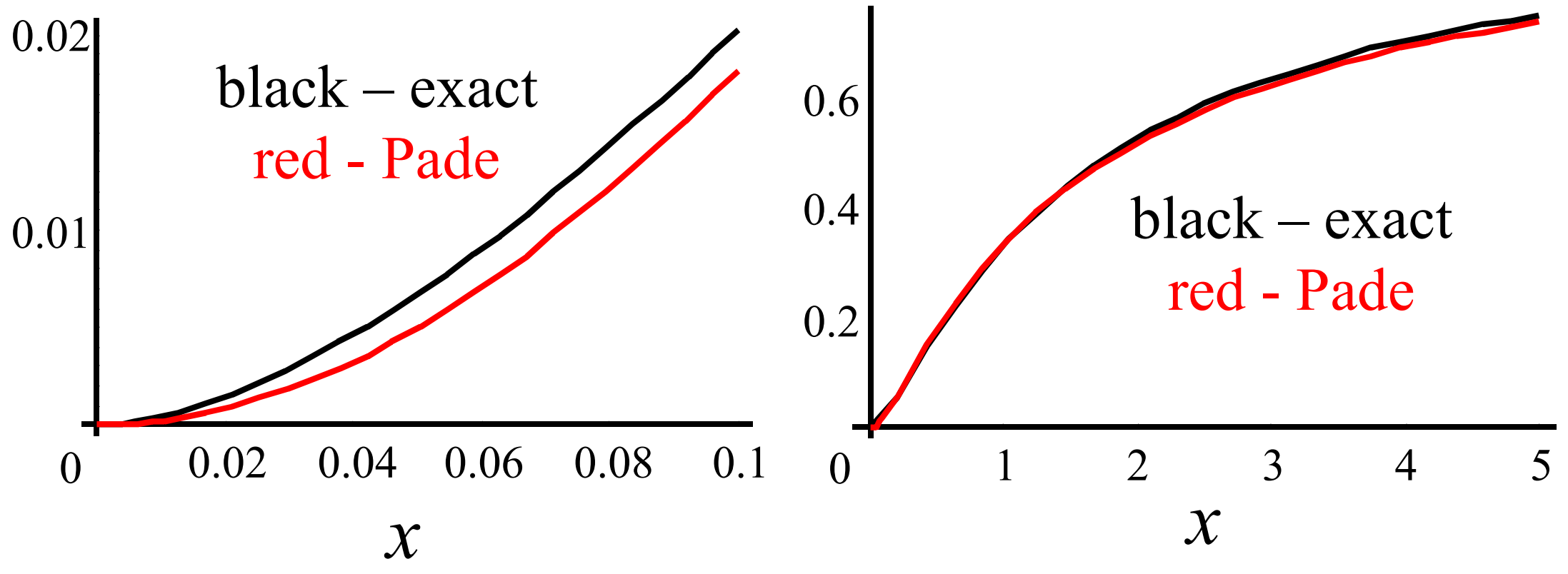

The excess of the energy $\Delta\Phi_{DS} = \int\limits_{\Sigma(z>0)} ds\,(\mathbf{P}_S \cdot \mathbf{n})\varphi_{DS}$ has been calculated using the surface charges potential $\varphi_{DS}(x,y,z) \approx \frac{(\sigma_S - P_S)r}{\varepsilon_0}\int\limits_0^\infty dk \frac{J_0\left(k\sqrt{x^2+y^2}\right)J_1(kr)}{\varepsilon_e k + \kappa\sqrt{k^2+R_d^{-2}}}\exp\left(-z\sqrt{k^2+R_d^{-2}}\right)$ . Using Zommerfeld formulae $\frac{\exp\left(-\sqrt{x^2+y^2+z^2}/R_d\right)}{\sqrt{x^2+y^2+z^2}} = \int\limits_0^\infty dk \frac{k\,J_0\left(k\sqrt{x^2+y^2}\right)}{\sqrt{k^2+R_d^{-2}}}\exp\left(-\sqrt{k^2+R_d^{-2}}\cdot|z|\right)$, one obtains the estimation $\varphi_{DS} \sim \frac{(\sigma_S - P_S)r^2}{\varepsilon_0(\kappa+\varepsilon_e)}\frac{\exp\left(-\sqrt{x^2+y^2+z^2}/R_d\right)}{\sqrt{x^2+y^2+z^2}}$ valid at $r \neq 0$ . Under these conditions we derived $\Delta\Phi_{DS}(r,l) \sim \frac{2P_S(\sigma_S - P_S)r^3}{1+(l/r\gamma)}$ . Under the condition $l >> r$ it could be neglected.